# RAG4Tickets: AI-Powered Ticket Resolution via Retrieval-Augmented Generation on JIRA and GitHub Data

**Mohammad Baqar**
(baqar22@gmail.com), Cisco Systems Inc, CA, USA

**Abstract**: Modern software teams frequently encounter delays in resolving recurring or related issues due to fragmented knowledge scattered across JIRA tickets, developer discussions, and GitHub pull requests (PRs). To address this challenge, we propose a **Retrieval-Augmented Generation (RAG)** framework that integrates **Sentence-Transformers** for semantic embeddings with **FAISS-based vector search** to deliver context-aware ticket resolution recommendations. The approach embeds historical JIRA tickets, user comments, and linked PR metadata to retrieve semantically similar past cases, which are then synthesized by a **Large Language Model (LLM)** into grounded and explainable resolution suggestions. The framework contributes a unified pipeline linking JIRA and GitHub data, an embedding and FAISS indexing strategy for heterogeneous software artifacts, and a resolution generation module guided by retrieved evidence. Experimental evaluation using precision, recall, resolution time reduction, and developer acceptance metrics shows that the proposed system significantly improves **resolution accuracy, fix quality, and knowledge reuse** in modern **DevOps** environments.

**Keywords:** Retrieval-Augmented Generation (RAG), Semantic Search, FAISS, Sentence Transformers, CodeBERT, Knowledge Reuse, Automated Ticket Resolution, Large Language Models (LLMs), JIRA, GitHub, DevOps Automation, Context-Aware Retrieval, Embedding Indexing, Explainable AI, Software Maintenance, Developer Productivity, Hybrid Retrieval, Data Drift, Hallucination Mitigation, AI-Augmented Triage

## 1. Introduction

Modern software development teams rely heavily on issue tracking systems such as JIRA and collaborative platforms like GitHub to manage feature requests, bug reports, and code changes. However, as projects scale, the volume of tickets, developer comments, and associated pull requests (PRs) grows exponentially, leading to information overload. Developers often spend significant time searching for past resolutions of similar issues, interpreting scattered conversations, and understanding linked code changes. Prior research has shown that machine learning techniques can aid in bug classification and triage [1], yet such approaches often fall short when handling the semantic variability in real-world bug reports. For instance, a bug describing “UI crash when toggling feature flags in React 19” might not be directly matched with an earlier issue phrased as “application freeze due to concurrent rendering,” even though the root cause overlaps, reflecting broader challenges in applying traditional supervised methods to complex, evolving enterprise data [2].

To address this gap, we propose a Retrieval-Augmented Generation (RAG) framework that combines semantic retrieval with context-aware language models for ticket resolution. RAG has been shown to reduce hallucinations and improve factual accuracy by grounding model outputs in retrieved evidence [3], making it well-suited for enterprise contexts where precision and reliability are critical. Our system leverages Sentence-Transformers to create embeddings of JIRA tickets, user comments, and PR descriptions, FAISS (Facebook AI Similarity Search) to perform efficient approximate nearest neighbor (ANN) search across a large corpus of tickets and code metadata, and a Large Language Model (LLM) decoder that synthesizes retrieved evidence into grounded, context-rich resolution suggestions. This approach demonstrates how resolution latency can be reduced, organizational knowledge reuse improved, and duplicate engineering effort minimized, while incorporating linked PR information to provide actionable code-change insights that guided past fixes.

### Contributions

- A unified architecture for ingesting and linking JIRA and GitHub data.

- An embedding and retrieval strategy tailored for heterogeneous artifacts (tickets, comments, PRs).
- A decision pipeline where retrieved evidence guides LLM-generated resolution suggestions.
- An evaluation using both IR metrics (Recall@k, MRR) and developer productivity measures (time-to-resolution, acceptance rate).
- A case study on recurring issues in a React 19 microservice migration, showcasing practical deployment.

## 2. Related Work

**AI in Ticket Classification and Resolution.**

The application of artificial intelligence to software maintenance has gained significant attention in recent years. Automated ticket classification systems have been developed to triage incoming bug reports and feature requests, often using machine learning models trained on historical issue data [1]. Such systems reduce human workload by automatically tagging issues with relevant categories or assigning them to appropriate developers. Deep learning approaches, including recurrent neural networks and transformer-based architectures, have demonstrated superior accuracy over traditional keyword-based methods by capturing semantic relationships between ticket descriptions and resolution categories [2]. Despite these advances, end-to-end automation of ticket resolution remains a challenging task, largely due to the fragmented nature of enterprise knowledge distributed across issue trackers, version control systems, and communication platforms.

**Retrieval-Augmented Generation (RAG) in NLP.**

Retrieval-Augmented Generation (RAG) has emerged as a powerful paradigm in natural language processing to address the limitations of large language models (LLMs) in generating contextually accurate responses [3]. Unlike traditional generation methods that rely solely on pre-trained parameters, RAG incorporates an external knowledge retrieval mechanism, allowing the model to ground its outputs in relevant evidence. Studies have shown that RAG reduces hallucinations, improves factual consistency, and adapts more effectively to domain-specific queries [4]. This approach has been applied to domains such as open-domain question answering, scientific document analysis, and customer support chatbots, where grounding in authoritative knowledge is critical [5]. For enterprise ticket resolution, RAG is particularly promising, as it can integrate dynamic repositories of historical tickets, user comments, and related code commits into the resolution process.

**FAISS and Vector Similarity Search in Large-Scale Retrieval.**

Efficient retrieval is a cornerstone of any RAG-based system. Facebook AI Similarity Search (FAISS) has become a widely adopted framework for high-dimensional vector similarity search, enabling scalable retrieval of semantically related documents in real time [6]. By leveraging approximate nearest neighbor (ANN) algorithms, FAISS can handle millions of embeddings with high recall while maintaining low latency [7]. This capability is crucial in enterprise contexts, where the corpus of bug reports, code commits, and documentation can grow rapidly over time. Recent works have combined FAISS with sentence-transformer embeddings to build domain-specific retrieval systems that outperform traditional lexical search engines such as Lucene and ElasticSearch [8]. These advances make FAISS particularly suitable for ticket resolution pipelines, where timely and accurate retrieval of past solutions directly impacts developer productivity.

**Prior Works Linking JIRA + GitHub Analysis.**

Several studies have explored integrating software development artifacts from multiple platforms to improve bug localization and resolution. Tools such as BugLocator [9] and DeepLoc [10] have shown that analyzing source code changes alongside issue reports can significantly enhance fault localization accuracy. More recent research has extended this integration to combine issue tracker data (e.g., JIRA) with version control histories (e.g., GitHub) to identify developer activities, code-review outcomes, and bug-fix patterns [11]. However, most of these works rely on supervised learning with static datasets and do not fully leverage semantic retrieval mechanisms. By combining JIRA ticket history, user comments, and linked GitHub pull requests into a unified RAG framework, our work addresses this gap, providing a dynamic and adaptive approach to ticket resolution grounded in historical evidence.

## 3. System Architecture

Our proposed framework integrates heterogeneous software artifacts into a Retrieval-Augmented Generation (RAG) pipeline for automated ticket resolution. The architecture is divided into distinct but interdependent modules—data ingestion, embedding, retrieval, and generation—allowing modular upgrades and scalability. By leveraging both semantic embeddings and LLM reasoning, the system ensures that recommendations are not only contextually relevant but also practically actionable [3].

### 3.1 Data Sources

The first stage involves structured ingestion of data from multiple development platforms. JIRA is the primary source of issue descriptions and metadata, GitHub contributes pull requests and associated code changes, and developer comments capture implicit knowledge such as stack traces and debugging hints. The linking between tickets and PRs ensures a holistic context that goes beyond surface-level text matching.

**Table 1: Example Data Sources and Extracted Fields**

| Source | Extracted Features | Example |
|---|---|---|
| JIRA Tickets | Title, description, priority, status, resolution | "UI crash on feature flag toggle" |
| User Comments | Discussions, error logs, patch hints | "Crash due to null pointer at render()" |
| GitHub PRs | Commit messages, diff summaries, review comments | "Fix: safeguard concurrent rendering" |
| Ticket–PR Links | Explicit issue keys in PR descriptions | "Fixes PROJECT-123 in commit #ab12cd" |

This integrated dataset forms the backbone of the pipeline, enabling cross-referencing of textual and code-level signals.

### 3.2 Embedding Layer

The embedding layer converts heterogeneous artifacts—textual (tickets, comments) and semi-structured (PR metadata, commit messages)—into dense vector representations. We utilize **Sentence-Transformers** such as *all-MiniLM-L6-v2* and *multi-qa-MPNet-base-dot-v1*, optimized for semantic similarity and cross-domain retrieval tasks [8]. For code-specific data, **CodeBERT** and **GraphCodeBERT** embeddings can be integrated to capture programming-language semantics, structural dependencies, and naming conventions.

To maintain retrieval precision, embeddings are **partitioned by artifact type** (tickets, comments, PRs), enabling targeted search spaces. This fine-grained separation allows the system to merge complementary evidence—e.g., a ticket's description may surface historical incidents, while PR embeddings highlight concrete implementation fixes. Embeddings are normalized and periodically refreshed to capture evolving terminology and codebases.

### 3.3 Retrieval Layer

All embeddings are indexed using **FAISS (Facebook AI Similarity Search)**, supporting large-scale **Approximate Nearest Neighbor (ANN)** retrieval across millions of vectors [14]. The index type—Flat, IVF, or HNSW—is selected based on corpus size and latency requirements.

**Workflow:**

A. A new ticket or issue is encoded into an embedding vector.
B. FAISS retrieves the **top-k semantically closest** tickets, PRs, and user comments.
C. Retrieved artifacts are re-ranked using cosine similarity and contextual overlap before being passed to the generation layer.

This approach ensures that suggestions are both **semantically aligned and contextually grounded**, reducing false positives in retrieval-heavy domains.

### 3.4 Generation Layer

The **Retrieval-Augmented Generation (RAG)** workflow synthesizes the final response by concatenating the top-ranked retrievals into a structured prompt for a **Large Language Model (LLM)**. The prompt template incorporates ticket metadata, relevant PR summaries, and historical resolutions. The LLM (e.g., GPT-4, Claude, or LLaMA 3 fine-tuned model) produces candidate solutions that include:

- Step-by-step resolution plans aligned with retrieved fixes.
- Hyperlinks to related PRs or commits for immediate traceability.
- Confidence scores or natural-language rationales derived from contextual matching.

Prompt optimization (e.g., token budgeting, truncation, and instruction tuning) ensures factual grounding and minimizes hallucination, enabling explainable and auditable recommendations.

### 3.5 Deployment Options

The modular design allows flexible integration within existing DevOps pipelines:

- **JIRA Integration:** An automated triage bot posts context-aware resolution suggestions within ticket comments.
- **GitHub Action:** When new PRs link to issues, historical resolutions are surfaced automatically for cross-reference.
- **CI/CD Integration:** The system detects and classifies recurring test failures, triggering automated RCA (Root Cause Analysis) suggestions.

Each mode supports **real-time monitoring, versioned embeddings, and rollback mechanisms** to ensure resilience. This adaptability facilitates adoption across organizations of varying maturity, ensuring measurable improvements in **MTTR (Mean Time to Resolution)** and engineering productivity.

**Table 2. System Architecture Components**

| Layer | Component | Tools / Models | Role in Pipeline |
|---|---|---|---|
| **Data Sources** | JIRA Tickets | JIRA API | Provides issue titles, descriptions, priorities, statuses, and resolutions. |
| | User Comments | JIRA Discussions, Logs | Captures developer discussions, stack traces, and hints for debugging. |
| | GitHub PRs | GitHub API | Supplies commit messages, code diffs, and review discussions. |
| | Ticket–PR Links | Issue Keys (e.g., PROJECT-123 in PR text) | Establishes connections between tickets and code changes. |
| **Embedding Layer** | Ticket Descriptions / Comments | Sentence-Transformers (all-MiniLM, MPNet) | Converts text into semantic dense vectors. |
| | PR Metadata (commits, diffs) | Sentence-Transformers / CodeBERT | Embeds structured code-related data. |
| **Retrieval Layer** | Vector Index | FAISS (Flat, HNSW, IVF) | Performs Approximate Nearest Neighbor (ANN) search to find similar tickets/PRs. |
| **Generation Layer** | Context Synthesis | RAG with LLM (e.g., GPT, LLaMA) | Combines retrieved context into grounded resolution suggestions. |
| | Evidence Linking | Retrieved Tickets + PRs | Cites past fixes and links to relevant GitHub commits. |
| **Deployment** | Resolution Bot in JIRA | JIRA Plugin / API | Suggests resolutions in ticket comments. |
| | GitHub Action | GitHub CI/CD | Suggests fixes in PRs linked to tickets. |
| | CI/CD Integration | Jenkins / GitHub Actions / GitLab CI | Auto-triage recurring failures during pipelines. |

### 3.6 Design Trade-offs and Rationale

The proposed RAG-based system was designed with an emphasis on scalability, explainability, and cost efficiency, driving several key architectural decisions. While managed vector databases such as Pinecone, Weaviate, or Milvus offer built-in scalability and distributed indexing, FAISS was chosen for its on-premise control, GPU acceleration, and customizable ANN indexing options (e.g., Flat, IVF, HNSW). This allows tighter integration with existing enterprise security and compliance frameworks, especially when handling sensitive ticket and code data.

In embedding selection, transformer-based models such as all-MiniLM-L6-v2 and multi-qa-MPNet-base-dot-v1 were preferred over heavy fine-tuned models due to their high semantic recall with low latency. Optional inclusion of CodeBERT ensures cross-modal understanding between textual descriptions and code diffs—critical for software ticket resolution tasks.

For the generation layer, instead of using a fine-tuned proprietary LLM, the architecture leverages Retrieval-Augmented Generation (RAG) to ensure contextual grounding and traceability. This design choice minimizes hallucination risk, reduces dependence on retraining, and allows modular upgrades (e.g., replacing FAISS or LLM independently).

Overall, this configuration provides an optimal balance between retrieval precision, computational efficiency, and system transparency, making it deployable in enterprise-grade environments without sacrificing explainability or maintainability.

## 4. Methodology

The methodology of our proposed RAG-based ticket resolution system follows a structured pipeline, transforming heterogeneous software artifacts into actionable knowledge. While the system architecture outlines the core building blocks, the methodology emphasizes the step-by-step execution, optimization strategies, and integration points.

### 4.1 Data Preprocessing

The raw data from JIRA and GitHub requires extensive preprocessing to ensure consistent quality. For tickets, we normalize fields by removing boilerplate text (e.g., “Steps to reproduce,” “Expected result”), resolving duplicates, and standardizing timestamps. User comments are cleaned using natural language preprocessing steps such as tokenization, lowercasing, and stopword removal, while preserving stack traces and error logs that often hold diagnostic value [12]. For GitHub pull requests (PRs), we extract commit messages, patch diffs, and reviewer discussions. To reduce noise, code diffs are summarized using AST-based (Abstract Syntax Tree) parsers to retain only function-level changes rather than line-by-line diffs [13].

### 4.2 Embedding Generation

The preprocessed artifacts are embedded into dense vector representations using **sentence-transformers** (e.g., *all-MiniLM-L6-v2*, *multi-qa-MPNet-base-dot-v1*) [8]. Tickets, comments, and PR metadata are encoded separately to preserve context-specific semantics. For PR diffs, we employ a hybrid strategy: natural language embeddings for commit messages and **code-aware embeddings** (CodeBERT, GraphCodeBERT) for actual code snippets [14]. This hybrid embedding strategy ensures that both natural language reasoning and structural code similarities are captured.

### 4.3 Index Construction with FAISS

Once embeddings are generated, they are stored in a **FAISS index** to enable scalable similarity search across millions of vectors [6]. Depending on dataset size and query latency requirements, we experiment with multiple FAISS index types:

- **Flat (brute-force):** Highest accuracy, slower retrieval, suitable for small datasets.
- **HNSW (Hierarchical Navigable Small World Graph):** Balances accuracy and speed, suitable for medium-to-large datasets.
- **IVF (Inverted File Index):** Optimized for very large corpora with trade-offs in recall.

We empirically evaluate these index structures, selecting **HNSW** for its balance of speed and recall in organizational-scale datasets.

### 4.4 Query Workflow

For a new incoming ticket, the workflow proceeds as follows:

1. The ticket is embedded into the same vector space as historical data.
2. The embedding is queried against the FAISS index to retrieve the **top-k most similar tickets and PRs**.
3. The retrieved evidence is concatenated and passed as context to a **Large Language Model (LLM)**, which generates a grounded resolution suggestion.
4. The LLM output includes candidate resolution steps, references to relevant PRs, and (optionally) a **confidence score** indicating the reliability of the suggestion.

This retrieval-generation cycle forms the core of the system's intelligence, ensuring that answers are both semantically relevant and context-rich [3].

### 4.5 Integration with Developer Workflows

The final step involves embedding the system seamlessly into existing developer workflows. We provide two main integration options:

- **JIRA Resolution Bot:** The system posts a suggested resolution directly in the ticket comments, allowing developers to accept, modify, or reject it.
- **GitHub Action Integration:** For PR-linked issues, the system provides inline comments with past relevant fixes and code-change patterns.

By positioning the AI assistant within the same tools developers already use, we minimize context switching and increase adoption [15].

### 4.6 Comparative Analysis of FAISS Index Structures

One critical design decision in our methodology involves selecting the appropriate FAISS index type. The choice depends on dataset size, desired accuracy, and latency requirements. Table 1 summarizes the trade-offs among the most commonly used FAISS indices.

**Table 3. Trade-offs between FAISS index structures**

| Index Type | Accuracy | Query Speed | Memory Usage | Best Use Case |
|---|---|---|---|---|
| Flat (Brute Force) | ★★★★★ (Exact match) | Slow (linear scan) | High | Small datasets (<100k vectors), experiments |
| HNSW (Hierarchical Navigable Small World) | ★★★★☆ (Near-exact) | Fast (logarithmic search) | Moderate | Medium-to-large datasets (100k–10M vectors) |
| IVF (Inverted File Index) | ★★★☆☆ (Approximate) | Very fast (sublinear) | Low | Very large datasets (>10M vectors), web-scale retrieval |

In practice, we found **HNSW** to provide the best balance between **latency and accuracy**, particularly for enterprise-scale datasets such as millions of JIRA tickets and PRs. The Flat index ensures perfect recall but is impractical at scale, while IVF scales efficiently but sacrifices accuracy in retrieving highly similar tickets.

By empirically comparing index types on a real-world dataset of ~1.2M tickets and PRs, we observed that HNSW reduced average query latency by **73%** compared to Flat, with only a **2% drop in recall**. This balance makes it suitable for integration into CI/CD workflows where both accuracy and responsiveness are critical [6].

## 5. Case Study: Ticket Resolution with React 19 Migration

To demonstrate the practical application of the proposed system, we consider a real-world case study involving the migration of a large-scale web application from React 18 to React 19. The migration introduced several recurring issues, particularly within the microservices architecture where multiple teams maintained interconnected components. A recurring challenge was related to deprecated lifecycle methods and state management inconsistencies during the migration, which resulted in multiple JIRA tickets being raised across different services. These tickets often included developer comments, stack traces, and references to GitHub pull requests (PRs) that contained partial fixes or experimental workarounds [16].

Using the Retrieval-Augmented Generation (RAG) pipeline, the system embedded historical tickets and PRs into a vector database indexed with FAISS. When a new migration-related issue was filed, the pipeline automatically transformed the ticket description into an embedding and searched for semantically similar records. In this scenario, the system successfully retrieved past tickets related to `useEffect` dependency changes, state batching improvements, and modifications in React's concurrent rendering model [17]. Additionally, PRs that contained refactoring strategies, such as replacing deprecated APIs with new concurrent-friendly hooks, were surfaced as context.

The candidate resolution was then generated by the large language model (LLM), which synthesized patterns from retrieved tickets and PR diffs. For example, when a new ticket described "UI freezes due to concurrent rendering mismatch," the system suggested a fix pattern based on prior migrations—rewriting affected components with `useTransition` to manage deferred updates. The generated suggestion was concrete enough to guide developers toward a solution while still requiring their judgment for context-specific adjustments [18].

A feedback loop was incorporated into the workflow, allowing developers to upvote or refine the candidate resolutions. This input not only improved the system's relevance ranking over time but also ensured trust and adoption among engineering teams. Over multiple iterations, developers reported reduced time in searching historical tickets manually and faster convergence toward correct fixes. This validated the hypothesis that AI-assisted retrieval and suggestion pipelines can significantly accelerate the resolution of recurring migration issues [19].

The evaluation of our RAG-based ticket resolution framework focuses on three complementary perspectives: **retrieval quality**, **generation performance**, and **business impact**. These perspectives provide both technical and organizational validation for the system.

### 5.1 Retrieval Metrics

We first assess retrieval effectiveness using **Recall@k** and **Mean Reciprocal Rank (MRR)**. Recall@k measures the fraction of relevant past tickets/PRs retrieved within the top *k* results, while MRR reflects how highly relevant items are ranked on average. In our experiments with approximately **15,000 historical JIRA tickets** and **6,500 GitHub PRs** collected from a large-scale React 19 migration project, the FAISS HNSW index achieved:

- Recall@5: **82%**
- Recall@10: **91%**
- MRR: **0.78**

This indicates that relevant prior resolutions were usually retrieved within the first five results, confirming the efficiency of FAISS in handling multi-source embeddings [14].

### 5.2 Generation Metrics

The output of the RAG pipeline was evaluated using **BLEU**, **ROUGE-L**, and a **factual consistency score** computed against ground-truth developer resolutions. On a test set of 500 unseen JIRA issues:

- BLEU: **0.47**
- ROUGE-L: **0.62**
- Factual Consistency: **84%**

The factual grounding score, measured using attribution checks against retrieved documents [20], highlights that the LLM was largely faithful to retrieved content. This minimized hallucinations, a common challenge in generative AI systems [21].

### 5.3 Business Metrics

From an organizational standpoint, the key benefits were captured through **mean resolution time reduction**, **human acceptance rate**, and **developer productivity uplift**. Across three agile teams adopting the RAG system:

- Average resolution time reduced from **18.5 hours** to **10.2 hours** (45% improvement).
- Human acceptance rate (i.e., developers directly adopting or lightly editing AI-suggested resolutions) was **68%**.
- Developer surveys indicated a **32% self-reported productivity uplift**, primarily due to reduced repetitive triaging work.

### 5.4 Experimental Dataset

The dataset comprised:

| Data Source | Quantity | Notes |
|---|---|---|
| JIRA Tickets | 15,000 | Includes titles, descriptions, comments |
| GitHub PRs | 6,500 | Includes commit messages, diffs, review comments |
| Linked Tickets–PRs | 3,200 | Explicit mappings via issue keys |

These results collectively demonstrate that the proposed RAG framework not only performs competitively on standard retrieval/generation benchmarks but also yields tangible business value when integrated into software engineering workflows.

## 6. Analysis, Limitations, and Engineering Implications

Our findings confirm that a **Retrieval-Augmented Generation (RAG)** framework can significantly accelerate ticket triage and enhance the reuse of organizational knowledge. However, several challenges must be addressed before achieving reliable, production-scale adoption. The first concerns **factual reliability**—while factual grounding achieved approximately 84%, occasional hallucinations were observed when the LLM extrapolated beyond retrieved evidence. This limitation aligns with broader observations in generative AI research on the tension between fluency and factuality [22], [23]. Future iterations could incorporate **retrieval-verification layers** or **cross-encoder re-ranking** to ensure that generated recommendations remain strictly supported by retrieved artifacts.

A second limitation involves **historical bias** and **data quality** within JIRA and GitHub sources. If earlier resolutions reflected suboptimal engineering practices or incomplete fixes, the system risks reinforcing these patterns. Similar concerns regarding bias propagation and fairness have been documented across machine learning systems [24]. Mitigation strategies include **temporal filtering**, **confidence weighting**, and **developer feedback loops** that allow human review to gradually refine model reliability over time. Dataset **drift** also poses a significant technical challenge. As frameworks evolve (e.g., React 18 → React 19 or new testing libraries), the semantic landscape of issues shifts, degrading retrieval precision. Regular **re-embedding cycles**, **incremental FAISS index refreshes**, and **continual fine-tuning** of sentence-transformer models are essential for maintaining relevance and stability [25].

From a deployment perspective, the case study revealed that developers placed greater trust in AI-generated resolutions when **transparent evidence**—such as retrieved PR links and rationale—was displayed alongside suggestions. This corroborates prior research indicating that **explainability and interpretability** directly correlate with user adoption in human–AI collaboration [26]. However, real-world deployment introduces operational costs: indexing tens of thousands of artifacts in FAISS requires optimized **HNSW or IVF configurations**, and repeated LLM inference adds **compute and latency overhead**. Emerging solutions, including **hybrid retrieval pipelines** (dense + sparse), **prompt caching**, and **edge inference** using quantized models, may offer a balance between performance and cost-efficiency [27]. Ultimately, while RAG-based systems show strong potential in automating software maintenance and ticket resolution, their long-term success depends on mitigating hallucination risk, ensuring model adaptability to evolving technologies, and embedding AI recommendations seamlessly into the developer workflow with transparency and scalability in mind.

**Table 4. Key Challenges and Mitigation Strategies in RAG-Based Ticket Resolution**

| Challenge | Description / Impact | Mitigation Strategy |
|---|---|---|
| **Hallucination & Factual Inconsistency** | LLM occasionally generates responses not grounded in retrieved evidence, leading to unreliable or misleading suggestions. | • Retrieval grounding through evidence-weighted decoding<br>• Post-generation factual verification<br>• Retrieval confidence scoring and entailment checks |
| **Historical Data Bias** | Legacy tickets and PRs may contain outdated or low-quality fixes, propagating technical debt or suboptimal patterns. | • Data provenance tracking and quality filtering<br>• Temporal relevance weighting<br>• Human-in-the-loop curation of training and retrieval datasets |
| **Dataset Drift (Framework Evolution)** | Changes in frameworks (e.g., React 18 → React 19) reduce relevance of older examples and embeddings. | • Periodic re-indexing and embedding refresh<br>• Time-aware similarity search<br>• Continuous fine-tuning using recent artifacts |
| **Explainability and Developer Trust** | Developers hesitate to adopt AI recommendations without transparent supporting evidence. | • Evidence-linked outputs (PRs, diffs, tickets)<br>• Interactive provenance visualization<br>• Explainable retrieval highlighting relevant context |
| **Scalability and Cost Overhead** | Indexing large-scale repositories and frequent LLM inference increase compute and operational costs. | • Hybrid retrieval (FAISS + sparse search)<br>• Query batching and response caching<br>• Lightweight local LLMs and modular architecture |
| **Knowledge Obsolescence** | Rapid evolution in tooling and practices makes older embeddings less useful. | • Adaptive learning pipelines with online embedding updates<br>• Model distillation for new domains<br>• Incremental fine-tuning driven by recent commits |

## 7. Conclusion and Future Work

This paper presented a **Retrieval-Augmented Generation (RAG)** framework that unifies JIRA tickets, developer discussions, and GitHub pull requests into an integrated pipeline for **AI-assisted ticket triage and resolution generation**. By leveraging **sentence-transformer embeddings** for semantic representation, **FAISS-based Approximate Nearest Neighbor (ANN)** search for large-scale retrieval, and **Large Language Model (LLM)-driven synthesis** for contextual reasoning, the proposed system demonstrates how retrieval and generation can be effectively fused to emulate human-like diagnostic and decision-making behavior. Our case study on a **React 19 microservice migration** illustrated the system's capability to recall semantically related historical fixes and produce **actionable, evidence-grounded recommendations**, while empirical evaluation confirmed measurable improvements in **retrieval precision, recall@k, and time-to-resolution metrics**. The results collectively validate the promise of **AI-augmented DevOps workflows** to mitigate cognitive overload, accelerate triage, and enhance organizational knowledge reuse across software delivery lifecycles.

Looking forward, several **technically promising extensions** arise. Future work will aim to improve **factual grounding** and **result traceability** by incorporating a **retrieval verification layer** and **reinforcement learning with human feedback (RLHF)** to penalize hallucinations and reward accurate reasoning. To support **enterprise-scale workloads** spanning millions of artifacts, we plan to investigate **hybrid retrieval architectures** that combine **dense FAISS indexing** with **sparse retrievers** such as BM25 or ColBERT for precision–recall tradeoff optimization. Another frontier lies in **continual and adaptive learning**, where embeddings and model weights evolve dynamically as new frameworks (e.g., React 20 or Next.js 15) and coding patterns emerge, thereby preventing **semantic drift** and **knowledge staleness**. In addition, future iterations may integrate **graph-based embeddings** to better represent ticket–PR–developer relationships and improve contextual linking. Finally, practical and ethical considerations—such as **auditability**, **explainability**, **data privacy**, and **inference cost optimization**—must remain integral to ensure the system's trustworthiness and sustainable deployment in **real-world CI/CD ecosystems**.

## Key Takeaways

This study demonstrates how **Retrieval-Augmented Generation (RAG)** can transform traditional ticket resolution by combining semantic retrieval with AI-driven reasoning. Using embeddings, FAISS, and large language models, the framework bridges the gap between unstructured organizational knowledge and automated resolution workflows. The approach not only accelerates developer decision-making but also establishes a foundation for **self-improving, explainable, and scalable DevOps automation**. As future systems evolve toward **autonomous software maintenance**, integrating retrieval intelligence with adaptive LLMs represents a significant step toward **self-healing and context-aware engineering environments**.